%% file: manuscript.tex
\RequirePackage{fix-cm}
\pdfoutput=1
\documentclass[twocolumn,epjc3]{svjour3}          
\usepackage[utf8]{inputenc}
\smartqed  
\usepackage{graphicx}
\usepackage{amssymb,amsmath}
\usepackage{hyperref}

\usepackage{array}
\newcolumntype{L}[1]{>{\raggedright\let\newline\\\arraybackslash\hspace{0pt}}m{#1}}
\newcolumntype{C}[1]{>{\centering\let\newline\\\arraybackslash\hspace{0pt}}m{#1}}
\newcolumntype{R}[1]{>{\raggedleft\let\newline\\\arraybackslash\hspace{0pt}}m{#1}}

\usepackage{xcolor}
\usepackage{input/journalabrv}
\usepackage[style=nature,
            doi=false,
            isbn=false,
            url=true,
            backref=false,
            hyperref=true,
            sorting=none,
            backend=biber]{biblatex}
\DeclareNameAlias{default}{first-last} 
\AtEveryBibitem{\clearfield{title}}
\AtEveryBibitem{\clearfield{url}}
\DeclareFieldFormat[article]{volume}{\textbf{#1}}
\DeclareFieldFormat[article]{journaltitle}{#1}
\DeclareFieldFormat[article]{pages}{#1}

\usepackage{lineno}

\DeclareSourcemap{
 \maps[datatype=bibtex,overwrite=true]{
  \map{
    \step[fieldsource=Collaboration, final=true]
    \step[fieldset=usera, origfieldval, final=true]
  }
  \map{
    \step[fieldsource=comment, final=true]
    \step[fieldset=userb, origfieldval, final=true]
  }
 }
}

\renewbibmacro*{author}{%
 \iffieldundef{usera}%
   {\printnames{author},}%
   {\printnames{author} (\printfield{usera}),}%
}%
\renewbibmacro*{issue+date}{%
  \setunit{}%
  \addspace
  \iffieldundef{userb}%
      {(\printfield{year})}
      {(\printfield{year})%
       \setunit{\addspace}%
       \printfield{userb}%
      }%
  \newunit%
}

\journalname{Eur. Phys. J. C}
\begin{document}

\title{Search for Dark Matter Annihilation in the Galactic Center with IceCube-79
}
\input{input/authors_25052015.tex}
\date{Received: date / Accepted: date}

\onecolumn
\maketitle

\twocolumn
\begin{abstract}
The Milky Way is expected to be embedded in a halo of dark matter particles,
with the highest density in the central region, and decreasing density
with the halo-centric radius.
Dark matter might be indirectly detectable at Earth through a flux of stable
particles generated in dark matter annihilations and peaked in the
direction of the Galactic Center.

We present a search for an excess flux of muon \mbox{(anti-)} neutrinos from dark matter
annihilation in the Galactic Center using the cubic-kilometer-sized IceCube
neutrino detector at the South Pole. There, the Galactic Center is always
seen above the horizon. Thus, new and dedicated veto
techniques against atmospheric muons are required to make the southern hemisphere accessible for IceCube.

We used 319.7 live-days of data from IceCube operating in its 79-string
configuration during 2010 and 2011. No neutrino excess was found and the
final result is compatible with the background. We present
upper limits on the self-annihilation cross-section, $\left<\sigma_\mathrm{A}
v\right>$, for WIMP masses ranging from 30\,GeV up to 10\,TeV, assuming cuspy (NFW) and
flat-cored (Burkert) dark matter halo profiles, reaching down to $\simeq 4 \cdot
10^{-24}$\,cm$^3$\,s$^{-1}$, and $\simeq 2.6 \cdot 10^{-23}$\,cm$^3$\,s$^{-1}$
for the $\nu\overline{\nu}$ channel, respectively.

\keywords{Dark Matter \and Galactic Center \and Indirect Search \and Neutrinos \and IceCube \and DeepCore}
\PACS{95.35.+d \and 98.70.Sa \and 98.35.-a}
\end{abstract}

\section{Introduction}

The first clear evidence for the existence of an invisible mass component in
the universe was Zwicky's observation of the dynamics of the Coma galaxy
cluster~\cite{1933AcHPh...6..110Z}. Subsequently, a broad range of cosmological
and astrophysical observations supported the existence of this dark matter
(DM) on various scales, from galaxy cluster scales down to galactic scales.
Measurements of galactic velocity profiles hint at invisible mass distributed
beyond the visible disks~\cite{1976ComAp...6..105R}. Galaxy cluster dynamics
exhibit a similar behavior~\cite{1996ApJ...462...32C}.

Further evidence for the existence of dark matter can be found in galaxy
cluster mergers like the Bullet
Cluster~\cite{1538-4357-496-1-L5,1538-4357-648-2-L109}.  Following a collision,
the interstellar and intergalactic gas components as seen in X-ray observations
are spatially separated from the reconstructed mass distribution.  Such a
separation strongly disfavors theories of modified gravity.

According to the current understanding of the formation and evolution of
large-scale structures, cold (non-relativistic), or warm dark matter is
preferred over hot (relativistic) dark matter. Otherwise, the formation of the
observed large-structures on time scales of the order of the age of the
universe would not have been
possible~\cite{2009MNRAS.398.1150B,Springel:2006vs,Ade:2013zuv}.

Though the nature of dark matter is largely unknown, some of its
properties may be deduced from the above-mentioned observations.  Analyses of
temperature fluctuations in the Cosmic Microwave Background (CMB) by the Planck
collaboration~\cite{Ade:2013zuv} yield the current best estimate for the total
content of DM in the universe: $\Omega_{\rm CDM} h^2 = 0.1199 \pm 0.0027$, with the
cold DM density parameter $\Omega_{\rm CDM}$, and $h=0.673 \pm 0.012$ being the
Hubble parameter divided by \mbox{$100\,\mathrm{km/s\,Mpc}$}.

Besides inference from gravitational interaction, particle DM may also be
detected indirectly. A weakly interacting massive particle (WIMP) at
roughly GeV-scale masses is a favorable class of DM; it naturally provides the
right order of magnitude for the thermal relic abundance of DM in the early
universe~\cite{Jungman:1995df}.
Examples of WIMPs are neutralinos in supersymmetric extensions of the Standard
Model~\cite{2010pdmo.book..142E}, or the lightest stable excitations in
Kaluza-Klein models~\cite{Hooper:2007qk}.

If DM decays, or (self-)annihilates, a flux of stable final-state messenger
particles, e.g. charged leptons, photons, and neutrinos, may be detected at
Earth, making DM experimentally accessible by indirect
searches (e.g.~\cite{Ellis1988403,1998APh.....9..137B,2005MPLA...20.1021B}).
The neutrino is an attractive messenger particle because it propagates without
absorption, and neutrino vacuum oscillations do not alter the energy and
direction information. Further, no fore- or background from astrophysical
objects has been identified yet.
Regions of increased DM density, like massive celestial objects, dwarf
galaxies, galactic halos, and the Galactic Center, provide targets to search
for an increased flux of neutrinos. Due to its proximity, the Galactic Center
is expected to yield the highest flux of annihilation products~\cite{Yuksel:2007ac}.
While most of these sources would appear as (nearly) point-like sources in the
sky, the Galactic Center is an extended source, and a signal from the Milky Way
halo would lead to a large-scale anisotropy in neutrino arrival directions
~\cite{2011PhRvD..84b2004A,Aartsen:2014hva}. With its 4$\pi$ acceptance, the
IceCube neutrino detector~\cite{doi:10.1146/annurev-nucl-102313-025321}, is
well-suited for DM searches from all of the above-mentioned sources.

In this paper we present the results from a search for a neutrino signal from
DM self-annihilation in the Galactic Center, targeting DM masses ranging from
$30~\mathrm{GeV}$ to $10~\mathrm{TeV}$. Due to the wide range of event
topologies associated with neutrinos from this energy range, two event
selections are motivated and presented. One event selection focuses on
the low-mass region ranging from 30\,GeV to 100\,GeV, accessible through the low-energy in-fill array DeepCore
(DC)~\cite{bib:DeepCore}, with the surrounding parts of IceCube used as veto.
The second event selection focuses on the mass range 100\,GeV--1\,TeV, but extends up to 10\,TeV. For this
selection a larger part of the IceCube detector is defined as fiducial volume.
Throughout this paper we denote the low-mass event selection as LE and the
high-mass selection as HE.

\section{Dark Matter Halos}

DM halos are considered to be gravitationally self-bound overdensities of DM particles, formed through
hierarchical merging of proto-halos from primordial density fluctuations~\cite{1974ApJ...187..425P}.
There is a tension between halo profile fits to DM overdensities in N--body
simulations, and fits to observational data (the cusp-core
problem)~\cite{deBlok:2009sp}. N--body simulations seem to favor cuspy halos,
while observations of e.g.  dwarf spheroidal galaxies imply a rather flat
central core region.  However, the inner part of the halo profile may depend on
the host halo mass~\cite{2003MNRAS.344.1237R}. Simple models of DM halos
describe the density by a smooth spherically-symmetric function of the
halo-centric radius, with the maximal density at the center, and a decreasing
density with increasing radius. One parametrization of such density profiles is
given by (modified from~\cite{1996MNRAS.278..488Z})

\vspace{-4mm}
\begin{equation}
    \label{eq:rho}
    \rho_{\mbox{\tiny DM}} (r) = {\rho_0 \over
        \left ( \delta + \frac{r}{r_s} \right )^\gamma \cdot
        \left (
            1 + \left ( \frac{r}{r_s} \right )^\alpha
        \right )^{ (\beta  - \gamma)/\alpha}
        },
\end{equation}
with the shape parameters $\alpha$, $\beta$, $\gamma$, the scale radius $r_s$,
and the mass density normalization $\rho_0$, which is usually determined from
the assumed local DM density, $\rho_{\mathrm{local}}$, in our solar system. We
introduced the parameter $\delta$ to allow for a central core if set to 1,
while $\delta=0$ describes a cuspy halo profile.

The parametrization of Eq.~(\ref{eq:rho}) is a combination of power laws,
where e.g. the power-law index $\gamma$ describe the inner slope, while
$\alpha$ and $\beta$ describe the outer slope. 
The halo-centric distance of this transition region depends on the scale radius
$r_s$. 

In view of the unresolved cusp-core problem, we present results for two
halo density profiles. The widely used Navarro-Frenk-White (NFW) profile represents cuspy
halos~\cite{1996ApJ...462..563N}, and is chosen for comparability among
different experimental results. The Burkert profile is chosen as representative
of flat-cored profiles~\cite{1995ApJ...447L..25B}.  Based on observation, the
latter profile is currently favored for the Milky Way~\cite{Nesti:2013uwa}.
Table~\ref{tab:halo_params} shows the parameter values for the two models used
in this work.

\begin{table}[h]
  \caption{DM Halo parameters used in this analysis. Taken from~\cite{Nesti:2013uwa}.
  }
  \label{tab:halo_params}
  \resizebox{\columnwidth}{!}{
  \begin{tabular}{ l r | c c }
   \hline 
  \multicolumn{2}{c|}{Parameter} &  NFW & Burkert   \\
  \hline
  $(\alpha,\: \beta,\:\gamma,\:\delta)$&                            & $(1,\:3,\:1,\:0)$         & $(2,\:3,\:1,\:1)$    \\[4pt] 
  $\rho_0$              &[$10^{7}M_{\odot}$/kpc\textsuperscript{3}] & $1.40^{+2.90}_{-0.93}$    & $4.13^{+6.2}_{-1.6}$ \\[4pt]
  $r_s$                 &[kpc]                                      & $16.1^{+17.0}_{-7.8}$     & $9.26^{+5.6}_{-4.2}$ \\[4pt]
  $\rho_{\mathrm{local}}$ &[GeV/cm\textsuperscript{3}]              & $0.471^{+0.048}_{-0.061}$ & $0.487^{+0.075}_{-0.088}$\\	      
 
  \end{tabular}
  }
\end{table}

\section{Neutrino Signal from Dark Matter Annihilation}

The flux of final state particles from annihilating dark matter depends on the DM mass
density squared, integrated along the line-of-sight (\textit{l.o.s.}) through the DM halo, and is
given by the $J_{\mathrm{a}}$-factor. Following e.g.~\cite{1998APh.....9..137B,Yuksel:2007ac},
the $J_{\mathrm{a}}$-factor is

\vspace{-4mm}
\begin{equation}
    J_{\mathrm{a}}(\Psi) = \int\limits_0^{l_{\rm{max}}} \mathrm{d}l ~
        \rho_{\mbox{\tiny DM}}^2 \left( \sqrt{ R^2_{\mbox{\tiny SC}} - 2l R_{\mbox{\tiny SC}} \cos{\Psi} +l^2}\right).
    \label{eq:lineOfSightAnn}
\end{equation}
Here, the density profile along the \textit{l.o.s.} is parametrized for a given
angle between the \textit{l.o.s.} and the direction of the center of the galaxy, $\Psi$.
The parameters are the radius of the solar circle,
$R_{\mbox{\tiny SC}} \approx 8.5~\mathrm{kpc}$, and the maximal distance from
the observer along the \textit{l.o.s.}, $l_{\rm{max}}$. The latter is

\begin{figure}[t]
    \centering
    \includegraphics[width=\columnwidth]{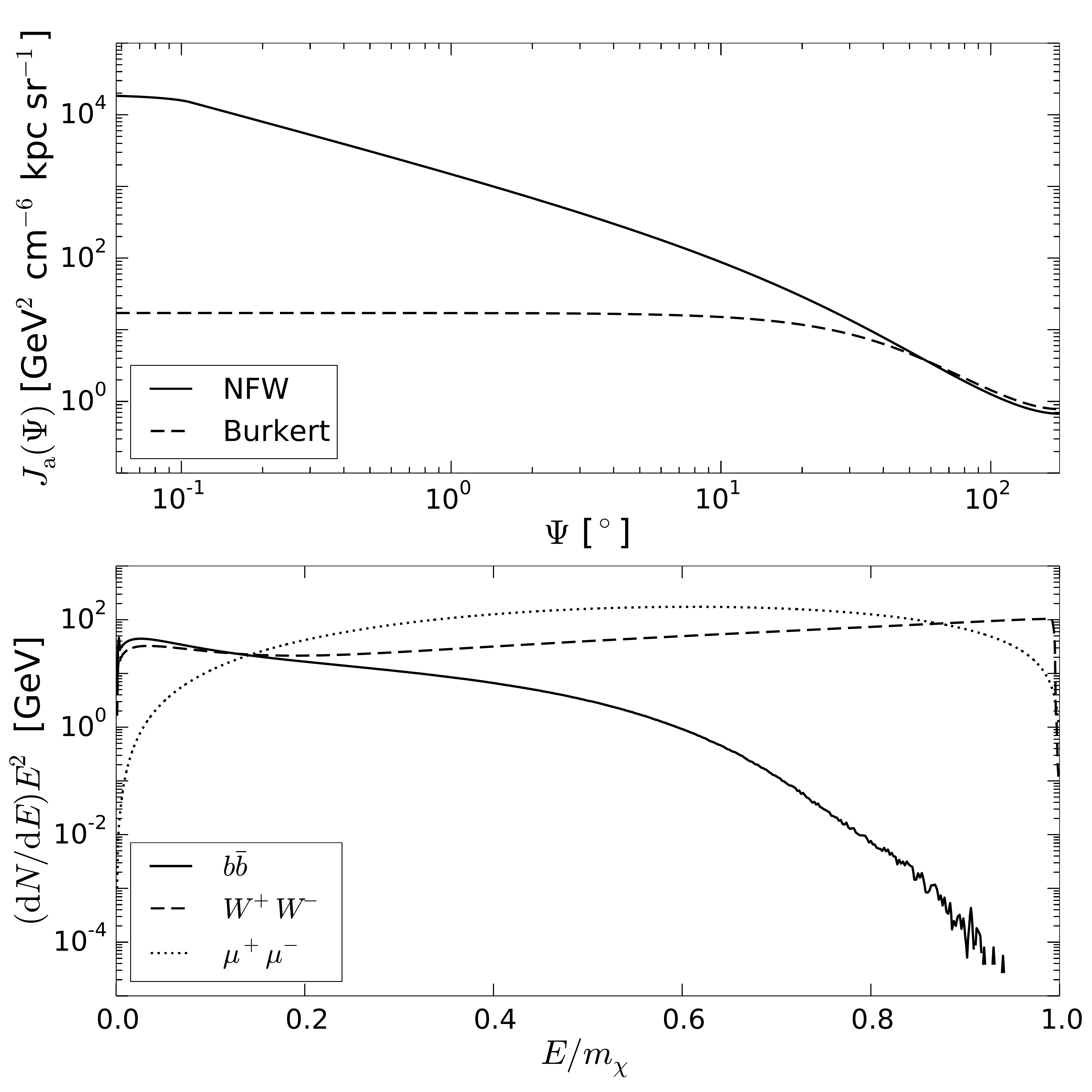}
    \caption[]{
        Top: Line-of-sight integral $J_{\mathrm{a}}(\Psi)$ of the NFW (solid) and Burkert 
(dashed) DM profile using parameters as given in Tab.~\ref{tab:halo_params}. 
        Bottom: Example of DM annihilation spectra generated with PYTHIA8~\cite{bib:Pythia8} for a 
WIMP mass of $m_\chi= 500$ GeV. 
        Three annihilation channels are shown: $b\bar{b}$ (solid), $W^+W^-$ (dashed), and $\mu^+\mu^-$ 
(dotted).
    }
    \label{fig:haloRhoJ}
\end{figure}

\vspace{-4mm}
\begin{equation}
    l_{\rm{max}} = \sqrt{R^2_{\mbox{\tiny MW}} -R^2_{\mbox{\tiny SC}} \sin^2{\Psi}  } + R_{\mbox{\tiny SC}} \cos{\Psi},
    \label{eq:lineOfSightAnnLMax}
\end{equation}
with the assumed radius of the Milky Way, $R_{\mbox{\tiny MW}} \approx
50~\mathrm{kpc}$. Typically, radii larger than the scale radius do not
contribute significantly to the total value of $J_{\mathrm{a}}$.
Figure~\ref{fig:haloRhoJ} (top panel) shows $J_{\mathrm{a}}$ for the NFW (solid line)
and Burkert (dashed line) profiles.

The final differential neutrino flux from DM annihilation, $\mathrm{d}\phi_\nu
/ \mathrm{d}E$, depends on the neutrino energy spectrum of the actual
annihilation channel.  The differential neutrino flux is

\vspace{-4mm}
\begin{equation}
    \frac{\mathrm{d}\phi_\nu}{\mathrm{d}E} = \frac{\left<\sigma_\mathrm{A} v\right>}{2}  ~ \frac{
1}{4\pi m_{\chi}^2} ~\frac{\mathrm{d}N_\nu}{\mathrm{d}E}~ J_{a}(\Psi),
    \label{eq:galaxyfluxanni}
\end{equation}
with $\left<\sigma_\mathrm{A} v\right>$ being the thermally averaged product of self-annihilation cross-section, $\sigma_\mathrm{A}$ and WIMP
velocity $v$. Further, $m_\chi$ is the WIMP mass, $\mathrm{d}N_\nu / \mathrm{d}E$ is the neutrino energy
spectrum per annihilating WIMP pair, and $J_{\mathrm{a}}$ is the
DM abundance along the \textit{l.o.s.}.

We consider several benchmark annihilation channels with 100\% branching ratios
($\chi \chi \rightarrow b \bar b$, $W^+ W^-$, $\mu^+ \mu^-$, $\tau^+ \tau^-$,$\nu \bar \nu$)
for the calculation of the neutrino energy spectrum $\mathrm{d}N_\nu /
\mathrm{d}E$. The resulting spectra bracket the realistic annihilation neutrino
energy spectra with a mixture of different annihilation branching ratios. We
generated a neutrino energy spectrum from annihilating DM for a particular WIMP
mass and annihilation channel using the PYTHIA8 (version 8.175) software
package \cite{bib:Pythia8}. Our PYTHIA8 simulation was set up to simulate a
generic resonance with an energy of twice the WIMP mass forming only the final
state particle pair (i.e. $b\overline{b}$, $W^+ W^-$, etc.) in question.
Subsequent processes like hadronization and decay were simulated using the
default PYTHIA8 implementations.  The generic resonance ensures an isotropic
decay of weak bosons, e.g. in the $W^+ W^-$ channel. Thus, the spin of the
annihilating WIMPs is not considered and we don't assume a specific WIMP model
like the lightest neutralino described by supersymmetric models
\cite{Jungman:1995df}.
If the WIMP is indeed the lightest neutralino, the spin (1/2) of the WIMP would
affect the generation of the neutrino energy spectra. The spin of such a WIMP
would lead to fully  transversely polarized $W$-bosons in the final state of
the annihilation process, thus altering the neutrino energy
spectrum~\cite{Barger:2007xf}. The differences in the differential neutrino
yield compared to the isotropically decaying $W$ bosons is about $\pm40\%$.
Examples of the neutrino energy spectra used here are shown in the bottom panel
of Fig.~\ref{fig:haloRhoJ} for the $b\overline{b}$, $W^+W^-$, and $\mu^+\mu^-$
annihilation channels.

In general, neutrinos are subject to neutrino oscillations on the way from the Galactic Center to 
the Earth. Due to the very long baseline, we assume a relative neutrino flavor ratio of 1:1:1 at Earth.

These simulations have the detector response folded in and are generated using
the ANIS event generator~\cite{Gazizov:2004va} modelling the neutrino-nucleon
charged and neutral current interactions via CTEQ5~\cite{Lai:1999wy} parton
distributions for neutrino and anti-neutrino interactions.

Finally, the neutrino energy distributions are used to weight generic simulated
neutrino data to DM annihilation signal. 

\section{The IceCube Neutrino Observatory}

The IceCube Neutrino Observatory, situated at the geographic South Pole,
consists of an in-ice detector array, IceCube, and a surface air shower
detector array, IceTop~\cite{2013NIMPA.700..188A}, dedicated to neutrino and
cosmic ray research, respectively.  IceCube~\cite{Achterberg2006155} is installed in
the glacial ice at depths between 1450\,m and 2450\,m below the surface,
instrumenting a total volume of one cubic kilometer.
IceCube detects neutrinos by optical detection of Cherenkov radiation induced
by secondary charged leptons which are produced in neutrino interactions in the
surrounding ice or the nearby bedrock. 

\begin{figure}[t]
\centering
\includegraphics[width=\columnwidth]{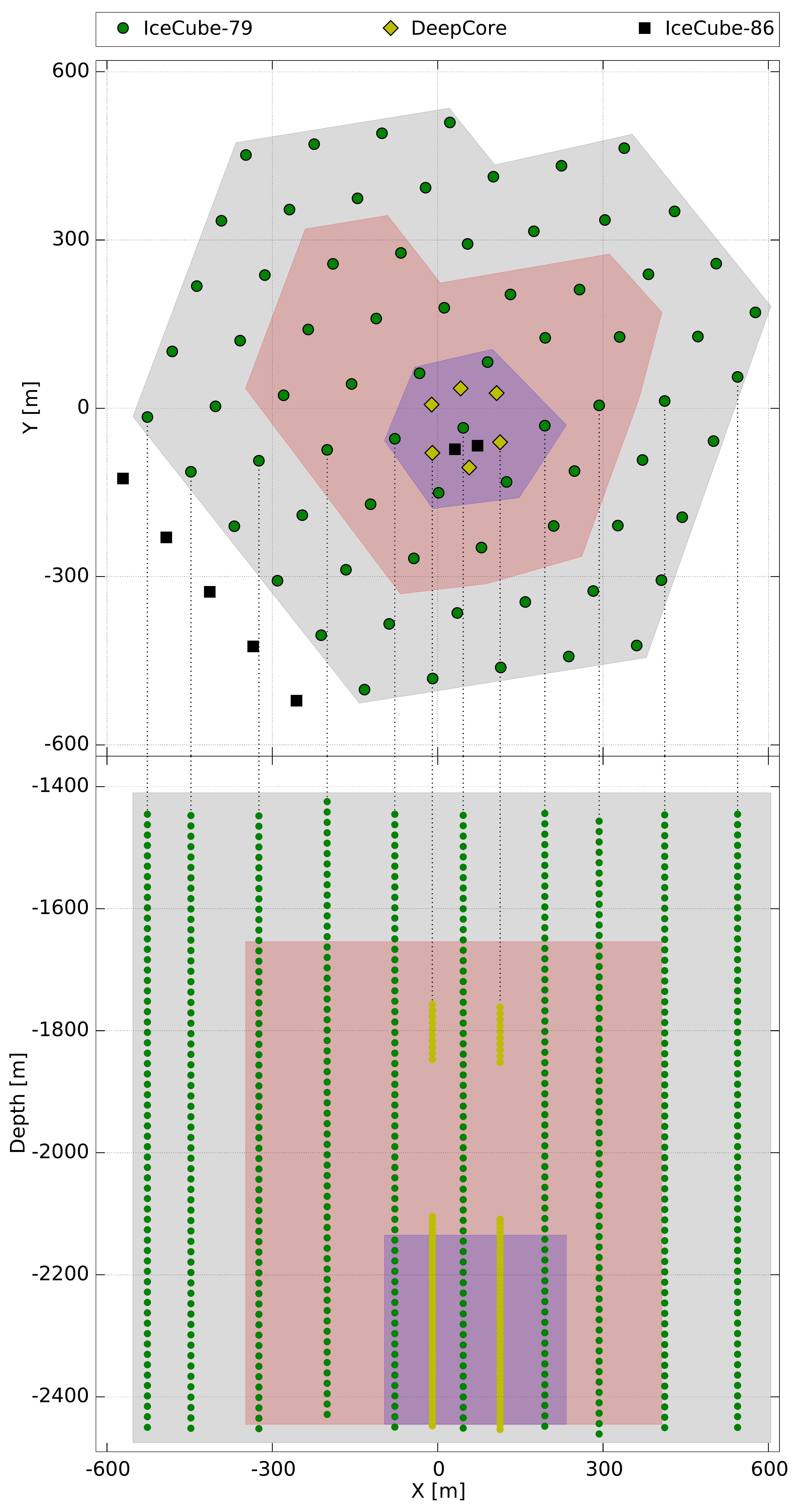}
\caption[]{
  A footprint (top) and side view (bottom) of IceCube in detector coordinates. 
  The green circles mark regular IceCube DOMs with 17\,m vertical spacing, while 
  the yellow diamonds mark DeepCore DOMs with about half the regular vertical DOM
  spacing and higher quantum efficiency PMTs. 
  The black squares mark DOMs which are part of the final 86-string configuration, but
  are not present in the here used 79-string configuration of IceCube.
  The purple shaded and red shaded areas illustrate the fiducial volumes used by the
  LE and HE event selections, respectively.
}
\label{fig:icGeo}
\end{figure}

Construction of the IceCube detector started in the Austral summer of 2004. In January 2010, 79 
detector strings with 60 digital optical modules (DOMs) on each string~\cite{Hanson2006214,
Achterberg2006155} were deployed. Each DOM contains a 25\,cm Hamamatsu photomultiplier tube
(PMT) and on-board electronics to readout and digitize the signal from the
PMT~\cite{bib:IceCubeDAQ}. In December 2010, the IceCube detector construction was completed. The final 
IceCube detector consists of 86 strings.
A schematic layout of the detector is shown in Fig.~\ref{fig:icGeo}. The 79
strings used in this analysis are marked by green and yellow markers within the
outer shaded grey area. The square markers denote the additional strings
constituting the completed IceCube detector.  Of the 79 strings used in this
analysis, 73 strings (green) have a horizontal spacing of
125\,m and a vertical spacing of 17\,m between DOMs. The six remaining strings
(yellow) are located near the central string of IceCube.
The DOMs on these strings are equipped with PMTs with a 30\% increased quantum
efficiency. Together with their nearest IceCube strings, these strings
constitute the inner detector, DeepCore~\cite{bib:DeepCore} (for IceCube-79).
The vertical distance between DOMs is reduced to 7\,m  (10\,m) for the bottom
50 (upper 10) DOMs.  The horizontal distance between strings in DeepCore is
less than 75\,m. 

These two densely-instrumented parts (see Fig.~\ref{fig:icGeo}) are separated by
a region with significantly reduced scattering and absorption lengths for
Cherenkov photons due to dust particles. It is located at a depth of about
2050\,m. 

For muon-neutrino events, the neutrino arrival directions are inferred
from the muon arrival direction. The latter is reconstructed using a likelihood
approach, based on the arrival times of photons at
DOMs~\cite{2004NIMPA.524..169A}. Thus, a good understanding of the absorption
and scattering of photons is necessary for direction reconstruction.
The clean glacial ice is a natural medium, built up over tens of thousands of
years. Thus, the optical properties exhibit a variation over the 1\,km depth of
the instrumented volume. 
A detailed description of the optical ice properties is given in~\cite{Aartsen:2013rt}.

\section{Event Selection}
The signal for this analysis is muons produced in charged-current neutrino
interactions.  These muons produce track-like event signatures in the detector,
which allow for a reconstruction of the arrival direction.

At the South Pole, the Galactic Center is always $29^\circ$ above the horizon.
Thus, neutrinos from the direction of the Galactic Center will appear as
down-going events in IceCube. The backgrounds for this analysis are therefore
down-going atmospheric muons and, at a lower rate, muons produced by
neutrinos originating from cosmic-ray showers in the atmosphere. 
The overwhelming majority of the 2500 Hz trigger rate is due to atmospheric muons.
The atmospheric neutrino background contributes to the trigger rate at
the 1\,mHz-level. This event class is an irreducible background, since in the
energy range of interest for this analysis the accompanying muon component of
the atmospheric shower is absorbed in the ice sheet above the detector.

The approach adopted here to reduce the background is to consider only neutrinos which
interact within the detector, and reject the background of penetrating (in-coming)
muons. In order to select events which appear to start within the detector we
developed several complementary veto techniques, exploiting differences in
timing and topology of background and signal events.

This work is based on two independently developed event selections, referred
to as low-energy (LE) and high-energy (HE) selections or samples.
There are two reasons for such energy-specific optimization. First, the efficiency
of these vetoes decreases rapidly with decreasing event energy, because
low-energy muons are able to traverse several string-layers without being
detected.  Second, the likelihood function (described in
Section~\ref{sec:Analysis}) does not use the event energy.

Though the individual event selections differ, the general selection techniques
and the analysis pipelines are very similar. First, an initial online selection
is performed at the South Pole. Second, a set of cuts on the event quality,
topology, and arrival directions is applied.  Third, a boosted decision tree
(BDT) is used to remove remaining background events, using the TMVA software
package~\cite{Hocker:2007ht}. The BDT is trained on a representative signal
assumption for each sample. Finally, a likelihood analysis is performed,
exploiting the different distributions of arrival directions of background
events and events originating in dark matter annihilations in the Galactic
Center. The differences between the two event selections are highlighted in
the following sections.

This analysis uses data collected with IceCube in its 79-string configuration
between May 31, 2010 and May 13, 2011 with a total live-time of 319.7 days of
stable high-quality data.  The LE sample contains 35,538 events, and the HE
sample contains 293,043 events.  4,706 events appear in both samples; about
13\% of the LE events are in the HE sample, and about 1.6\% of the HE events
are in the LE sample.

\subsection{Low-Energy Event Selection}\label{sec:LEEventSelection}

The LE event selection considers events from the DeepCore
online-filter~\cite{bib:DeepCore}, and is optimized for low-mass WIMPs below
100\,GeV, and thus uses the bottom part of the densely-instrumented DeepCore
sub-array as fiducial volume. The remaining instrumented IceCube volume as well
as the two bottom DOM layers are used as a veto.  The fiducial volume is
illustrated in Fig.~\ref{fig:icGeo}, and corresponds to roughly 27\,Mton of
ice.

The LE selection cuts are based on experience from the IceCube-79 Solar WIMP
analysis~\cite{bib:IC79SolarWIMP}, which used DeepCore for the first time in
low-mass WIMP searches. The signal used for BDT training are events that are
fully contained in the fiducial volume, and originate in annihilations of
65\,GeV WIMPs to $b \bar b$-pairs in the NFW halo. The search window for the LE
analysis extends to $\pm 30^\circ$ in right ascension ($\alpha$) with respect
to the Galactic Center, while the declination ($\delta$) width is asymmetric
and extends from $-39^\circ$ to $-9^\circ$.

The LE sample data rate at the analysis level is 1.4\,mHz.

\subsection{High-Energy Event Selection}\label{sec:HEEventSelection}

The IceCube array has a trigger energy threshold of about $\simeq $ 100\,GeV.
Therefore, the search for WIMPs in the mass range above a few hundred GeV
benefits from the large volume of IceCube in addition to DeepCore at the cost
of a decreasing veto efficiency. The HE selection considers events from the
dedicated Galactic Center online-filter and the DeepCore online-filter.

The veto for the HE event selection is defined by the upper 12 DOM layers and
the two outer string layers, which roughly corresponds to 200\,m and 125\,m of
instrumented distance, respectively. The fiducial volume for the HE selection
is shown in Fig.~\ref{fig:icGeo}. The signal assumed for BDT training are events
that are events which start in the fiducial volume, and originate in
annihilations of 600\,GeV WIMPs to $W^+W^-$-pairs in the NFW halo.
The search window around the Galactic Center is given by $\pm
15^\circ$ in both declination and right ascension.

The HE sample data rate at the analysis level is 10\,mHz.

\section{Analysis Method and Sensitivity}\label{sec:Analysis}

A maximum likelihood analysis is performed independently on each event
selection for a number of different WIMP masses ranging from 30\,GeV to
10\,TeV, assuming a 100\% branching ratio for each tested annihilation channel.

Considering the large number of events in the two final samples, a binned
likelihood method was chosen.  To reduce the number of bins, event
arrival directions were only considered in a search window around the Galactic
Center.  The search window shape and size differ slightly between the two event
samples as defined in the two previous sections~\ref{sec:LEEventSelection}
and~\ref{sec:HEEventSelection} as well as the background estimations.  Due to
these differences the likelihood analysis performed on the LE and HE selections
is not identical.  The LE likelihood has the more complicated form and is
defined as a function of signal fraction, $\xi$, in the following way:

\vspace{-4mm}
\begin{equation}
L(\xi) = \binom{N}{n}p^n(1-p)^{N-n}\prod^n_{i=1}f(\boldsymbol{X}_i,|\xi)
\label{eq:Likelihood}
\end{equation}
where the binomial factor in front of the product accounts for the probability
of observing $n$ events in the search window given $N$ total events in the
event selection, and the shape term, i.e the direction, $\boldsymbol{X} =
(\delta, \alpha)$, of the events is accounted for by the term
$f(\boldsymbol{X}_i,|\xi)$.  The binomial probability was chosen in favor of a
Poisson probability since the search window covers a non negligible fraction of
the declination band.  The probability of an event to fall in the search window
is defined as $p = \pi_\mathrm{s}\xi+\pi_\mathrm{bg}(1-\xi)$, where
$\pi_\mathrm{s}$ and $\pi_\mathrm{bg}$ are the probability for a signal or a
background event, respectively, to fall in the search window.  Note that the
signal originating in a dark matter halo is an extreme case of an extended
source as it is present in the whole sky.  Any background estimation based on
data will be contaminated by signal.  $\pi_\mathrm{bg}$ is determined from the
relation between the size of the search window and the size of the background
estimation region, while $\pi_\mathrm{s}$ is determined from simulation. 

\begin{figure}[t]
	\centering
	\includegraphics[width=0.48\textwidth]{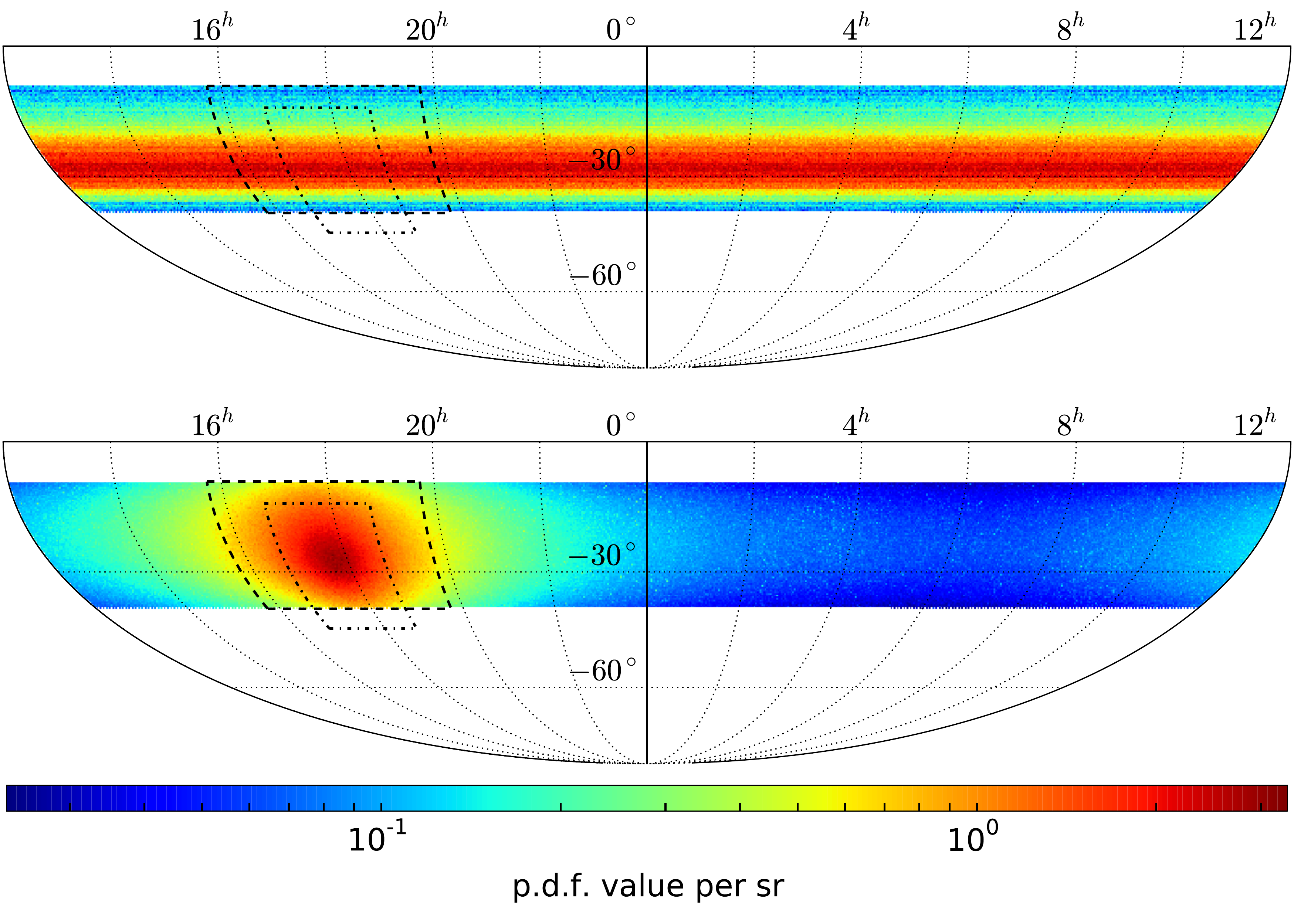}
	\caption[]{
		Example skymaps of the background (top) and signal (bottom) pdfs in equatorial coordinates 
for the LE event selection, for 100\,GeV WIMPs annihilating into $W^+W^-$. The LE search window is marked 
with a dashed line. For illustration the search window for the HE event selection is marked with a dash-dotted 
line.}
	\label{fig:skymapPDF}
\end{figure}

The directional probability density function (pdf), $f(\boldsymbol{X}|\xi)$, in
Eq.~(\ref{eq:Likelihood}) is constructed from binned expectations of event
directions for background and signal. Figure~\ref{fig:skymapPDF} shows examples
of these binned expectations.  The software package HEAL-Pix~\cite{bib:healpix}
was used to ensure equal-area bins on the sphere for these two-dimensional
pdfs.
To determine the signal pdfs, IceCube neutrino simulations were used, weighting events
according to Eq.~(\ref{eq:galaxyfluxanni}) and the corresponding DM annihilation spectrum.
Background pdfs were created by scrambling the right ascension of experimental data events
in the final event selections.

The same reasoning regarding signal contamination, as stated above, applies to the directional pdf, 
$f(\boldsymbol{X}|\xi)$.
In effect, the expected arrival directions of background events will depend on the signal strength. 
This needs to be accounted for in the directional pdf, $f(\boldsymbol{X}|\xi)$ in the likelihood, as well as
in the background simulation during the construction of confidence intervals. 
The directional pdf is defined as

\vspace{-4mm}
\begin{multline}\label{eq:PDF}
f(\boldsymbol{X}|\xi) = wf_\mathrm{s}(\boldsymbol{X})+\\
(1-w)[(1+w)f_{\mathrm{bg}}(\boldsymbol{X})-wf_{\mathrm{sc}}(\boldsymbol{X})]
\end{multline}
where $f_\mathrm{s}$ and $f_\mathrm{bg}$ are the signal and background directional pdfs, respectively, 
$f_{\mathrm{sc}}$ is a pdf describing the signal scrambled in right ascension. 
The signal fraction inside the search window is
\begin{equation}\label{eq:SignalFraction}
w = \frac{\pi_\mathrm{s}\xi}{\pi_\mathrm{s}\xi+\pi_\mathrm{bg}(1-\xi)}.
\end{equation}

A different approach is used for the HE analysis. The background estimation
is performed on off-source data, excluding all events within $\pm 30^\circ$ of
the Galactic Center. Therefore, any signal contamination of the background estimate is ignored, 
and the likelihood function from Eq.~(\ref{eq:Likelihood}) simplifies to:

\vspace{-4mm}
\begin{equation}\label{eq:Likelihood2}
	L(n_\mathrm{s}) = \frac{ (n_{\text{bg}}+ n_\mathrm{s})^{n} }{ n!
	}e^{-(n_\text{{bg}}+ n_\mathrm{s})} \prod_i^{n}
	f(\boldsymbol{X}_i , n_\mathrm{bg}|n_\mathrm{s}),
\end{equation}
where $n_\mathrm{s}$ is the number of signal events and $n_\mathrm{bg}$ is the
expected number of background events in the search window which is given by the
number of off-source events multiplied by the ratio of the on-source and
off-source region sizes.  The directional pdf consequently becomes:

\vspace{-4mm}
\begin{equation}\label{eq:PDF2}
f(\boldsymbol{X}, n_\mathrm{bg}|n_\mathrm{s}) =
\frac{n_\mathrm{s}}{n_\mathrm{bg}+n_\mathrm{s}}f_\mathrm{s}(\boldsymbol{X})+\frac{n_\mathrm{bg}}{n_\mathrm{bg} +n_\mathrm{s}}f_{\mathrm{bg}}(\boldsymbol{X})
\end{equation}

All confidence intervals are constructed using the prescription by Feldman and
Cousins~\cite{bib:FeldmanCousins}. The sensitivity is defined as the median
upper limit on the number of signal events at 90\% confidence level.

\begin{figure}[t]
    \centering
    \includegraphics[width=0.5\textwidth]{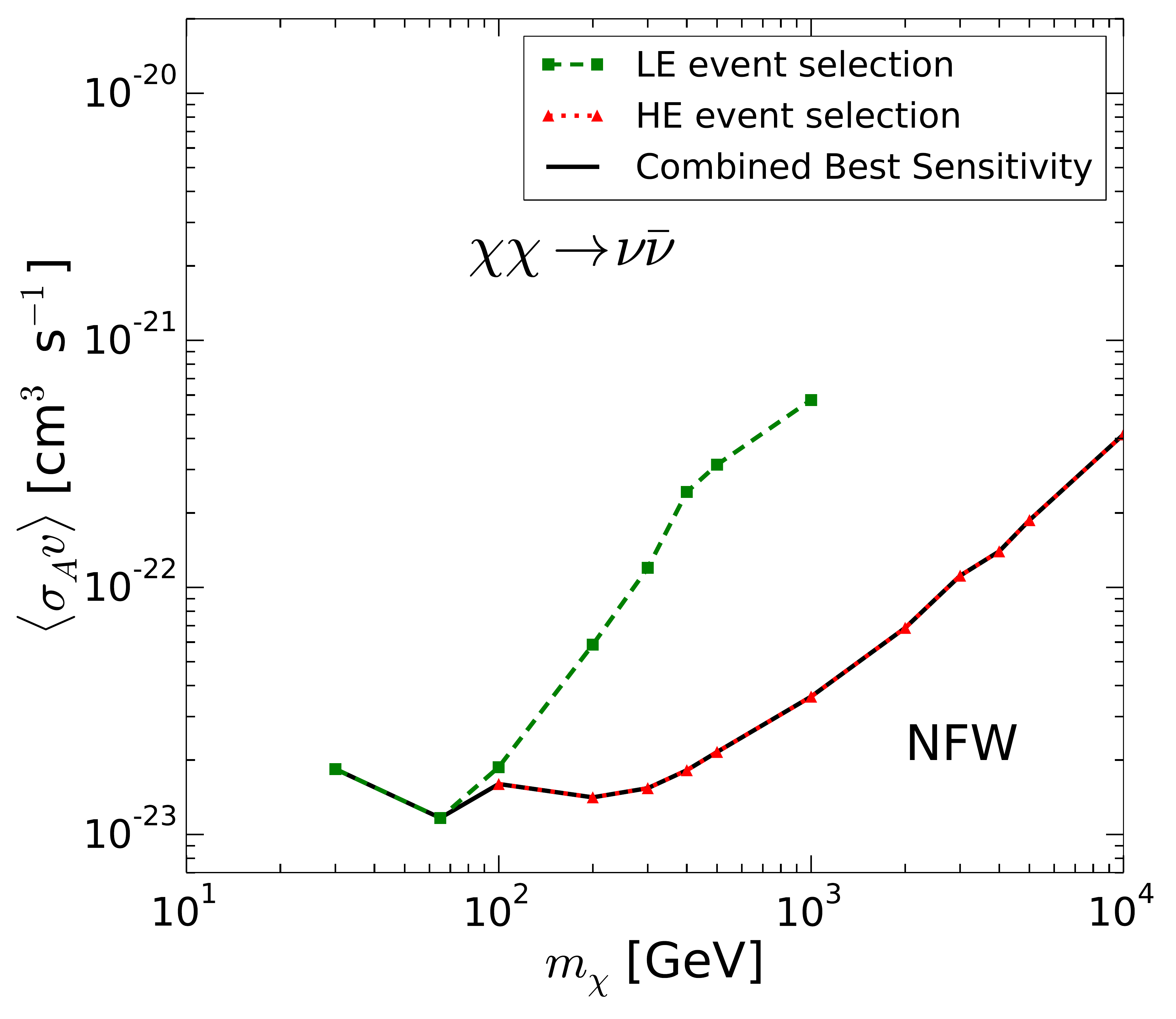}
    \caption[]{Median upper limits, i.e. sensitivities, at 90\% C.L. (w/o systematics) for the two
    event selections assuming WIMP annihilation to $\nu \bar{\nu}$ for the NFW DM profile. The black solid line
    shows the combined best sensitivity for this particular annihilation channel and DM 
	profile, considering both event selections. At 200\,GeV the HE selection
	yields a slightly better sensitivity and is thus used here.}
    \label{fig:sens-sel-comp}
\end{figure}

The final limits for each WIMP mass and annihilation channel are obtained from
the sample (LE or HE) which gives the best sensitivity (w/o systematics), i.e
the lowest median upper limit, for the particular WIMP mass and annihilation
channel. Thus, the cross-over point in the WIMP mass between the two event
samples depends on the DM halo model and WIMP annihilation
channel. This procedure circumvents the necessity to deal with the small
overlap of both samples.  Figure~\ref{fig:sens-sel-comp} illustrates how the
two event samples contribute to the best sensitivity at different WIMP
masses, in this case for WIMP annihilation to $\nu \bar{\nu}$. Each sample has a
WIMP mass range where it outperforms the other.

\begin{figure}[t]
    \centering
    \includegraphics[width=0.5\textwidth]{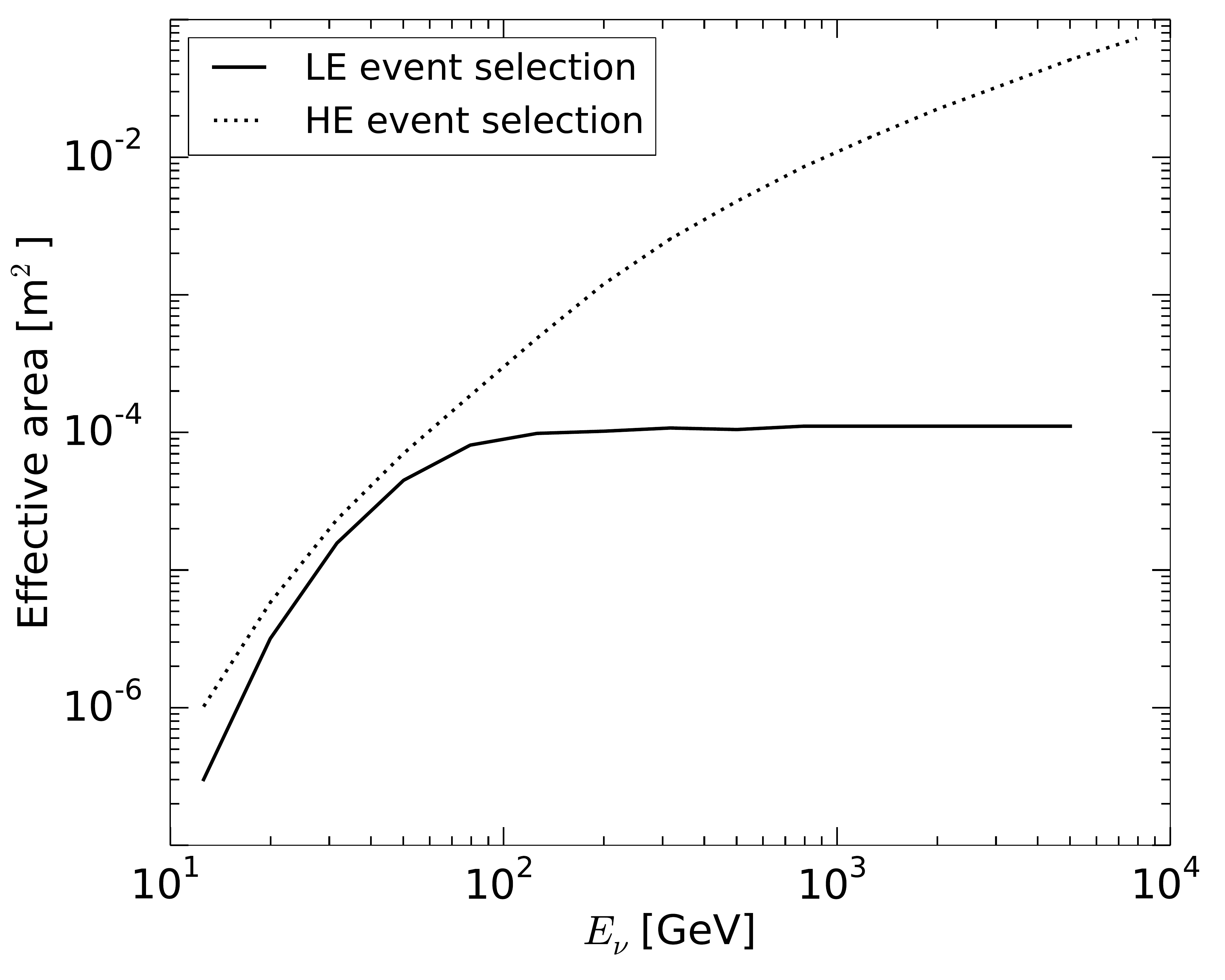}
    \caption[]{
        Neutrino effective areas as a function of energy for the two event selections. 
        The effective areas for the LE and HE selection are shown as solid, and dotted lines, respectively. 
        Although the HE effective area is bigger than the LE effective area at low 
        energies, the higher background contamination at low energies in the HE selection makes it less
        efficient.
    }
    \label{fig:effective_area}
\end{figure}

Figure~\ref{fig:effective_area} shows the neutrino effective area for the two
event selections.  Even though the effective area for the LE event selection is
smaller than that of the HE event selection, the sensitivity to the number of
signal events for low-mass WIMPs is better, as can be seen in
Fig.~\ref{fig:sens-sel-comp}. The reasons are the larger on-source region for
the LE event selection, and a lower background event rate due to higher veto
efficiency.

In order to avoid confirmation bias throughout the development of the analysis,
blindness with respect to the right-ascension information was imposed by
scrambling the right-ascension information of the experimental data. 
Only the declination information of the events was used for cut development.

Following the optimization of the event selections, meaning all cuts are fixed,
and the choice of the event selection for each channel, halo, and WIMP mass
(based on best sensitivity), the right ascension information was unblinded.

\section{Discussion of Uncertainties}

The uncertainties relevant for this analysis can be categorized into two classes:
\begin{itemize}
    \item Detector systematic uncertainties impacting the signal efficiency
    \item Astrophysical uncertainties (choice of halo model, model-specific parameter)
\end{itemize}
The former are incorporated into the calculated limits, while the latter are
studied to estimate model uncertainties. Both classes are discussed in the
following sections.

\subsection{Detector Systematics}
The uncertainties in the signal efficiency are mainly governed by the uncertainties in the optical efficiency
of the DOMs and the optical properties of the glacial ice, manifested in the absorption and 
scattering length. 
To determine the effects on $\langle \sigma_\mathrm{A} v \rangle$ due to the mentioned uncertainties, the
event selections and analysis were applied to sets of simulated data where the optical properties of the DOMs 
and the ice were changed. 
The optical efficiency of the DOMs was varied by $\pm10$\%. The same was done for the absorption and 
scattering lengths of the ice.
The resulting uncertainties on $\langle \sigma_\mathrm{A} v \rangle$ generally lie in the range 10\%~--~20\% 
except for the lowest neutrino energies where they reach up to $\approx 70\%$. 
This is due to threshold effects where events with just enough hit DOMs to trigger the
detector would fail to do so with increased absorption and scattering or decreased optical efficiency. 

The above-described systematic uncertainties are included into the limits by
degrading the baseline results by the relative variation of the detector
uncertainties with respect to the baseline, as stated above.

\subsection{Astrophysical Uncertainties}
\begin{figure}[t]
    \centering
    \includegraphics[width=0.5\textwidth]{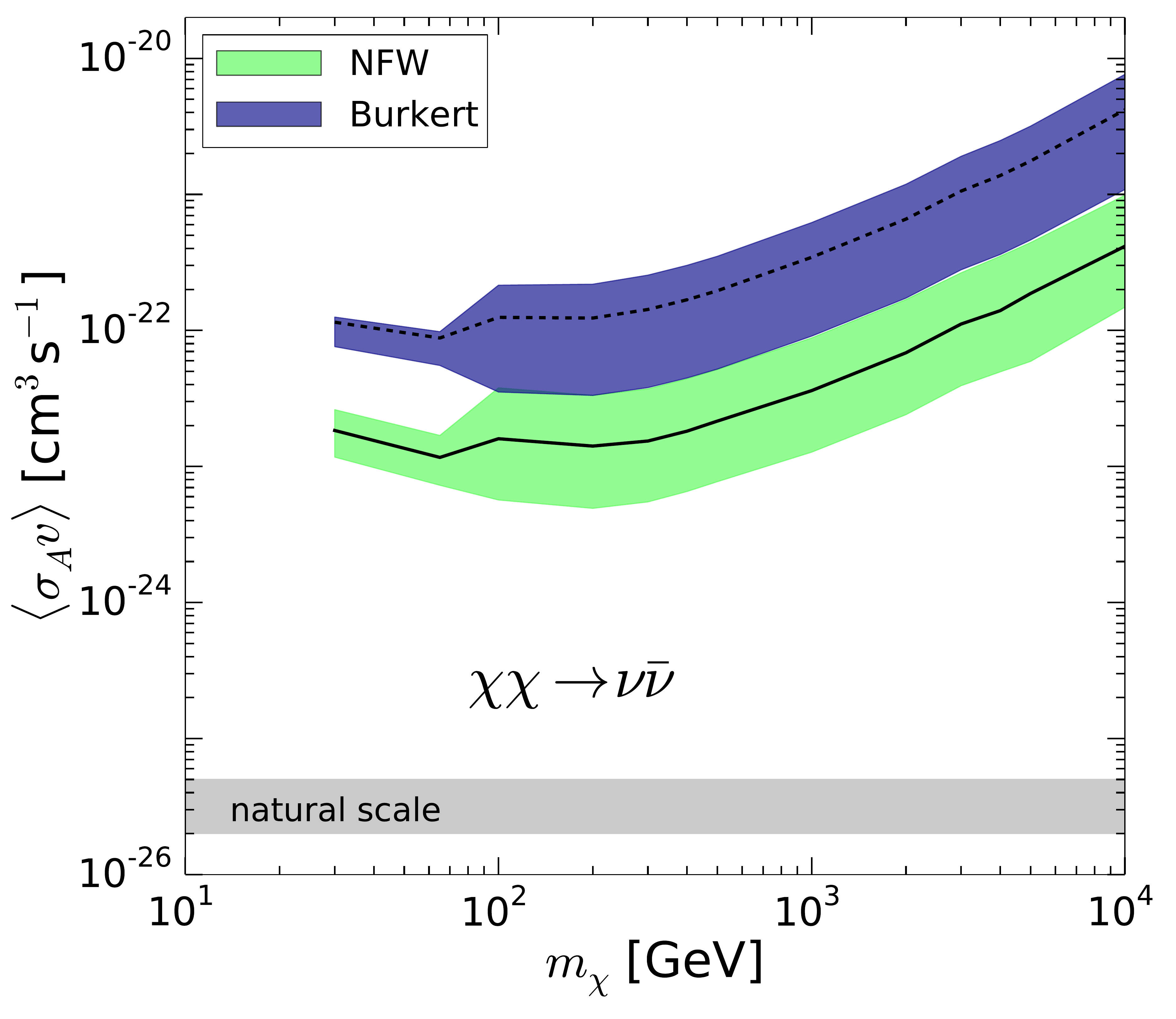}
    \caption[]{Uncertainties on the sensitivity due to the DM halo model parameter uncertainties for the NFW 
	(dashed line, blue band) and Burkert (solid line, red band) DM profile, assuming WIMP annihilation to 
	$\nu\overline{\nu}$. The reduced width of the bands below 100\,GeV is caused by different on-source
	regions, and thus differences in the integration of the J-factors. The dip below 100\,GeV is caused
	by the under-fluctuation in the LE sample.
	}
    \label{fig:sens-halo-par-unc}
\end{figure}

The astrophysical uncertainty is studied by using two different halo profiles,
and also by varying the parameters $\rho_{\mathrm{local}}$ and $r_{\rm s}$ within
the uncertainties stated in~\cite{Nesti:2013uwa}, and summarized in
Tab.~\ref{tab:halo_params}. Figure~\ref{fig:sens-halo-par-unc} compares the sensitivity for WIMPs 
annihilating to $\nu\bar\nu$-pairs for both profiles. 
The bands depict the variation of the sensitivity within each
profile that arises from varying the profile parameters within the given uncertainty.

The relative variation of $\langle \sigma_\mathrm{A} v \rangle$ due to the halo
profile parameter uncertainties was estimated to be 60\%~--~100\% for the LE
likelihood analysis and 60\%~--~200\% for the HE likelihood analysis, and is
shown in Fig.~\ref{fig:sens-halo-par-unc}. However, we refrain from including
uncertainties on the halo profile parameters into the limits.

\section{Results}

Table~\ref{tab:resultN} shows the number of events for the two unblinded event
selections. The quantities $n_\text{obs}$ and $n_\text{bg}$ are the number of
measured on-source events and expected background events in the search windows,
respectively. A $2\sigma$ under-fluctuation of experimental data events was
observed for the LE event selection. A systematic origin of this
under-fluctuation due to an uneven right-ascension exposure was excluded.

\begin{figure}[t]
    \centering
    \includegraphics[width=0.5\textwidth]{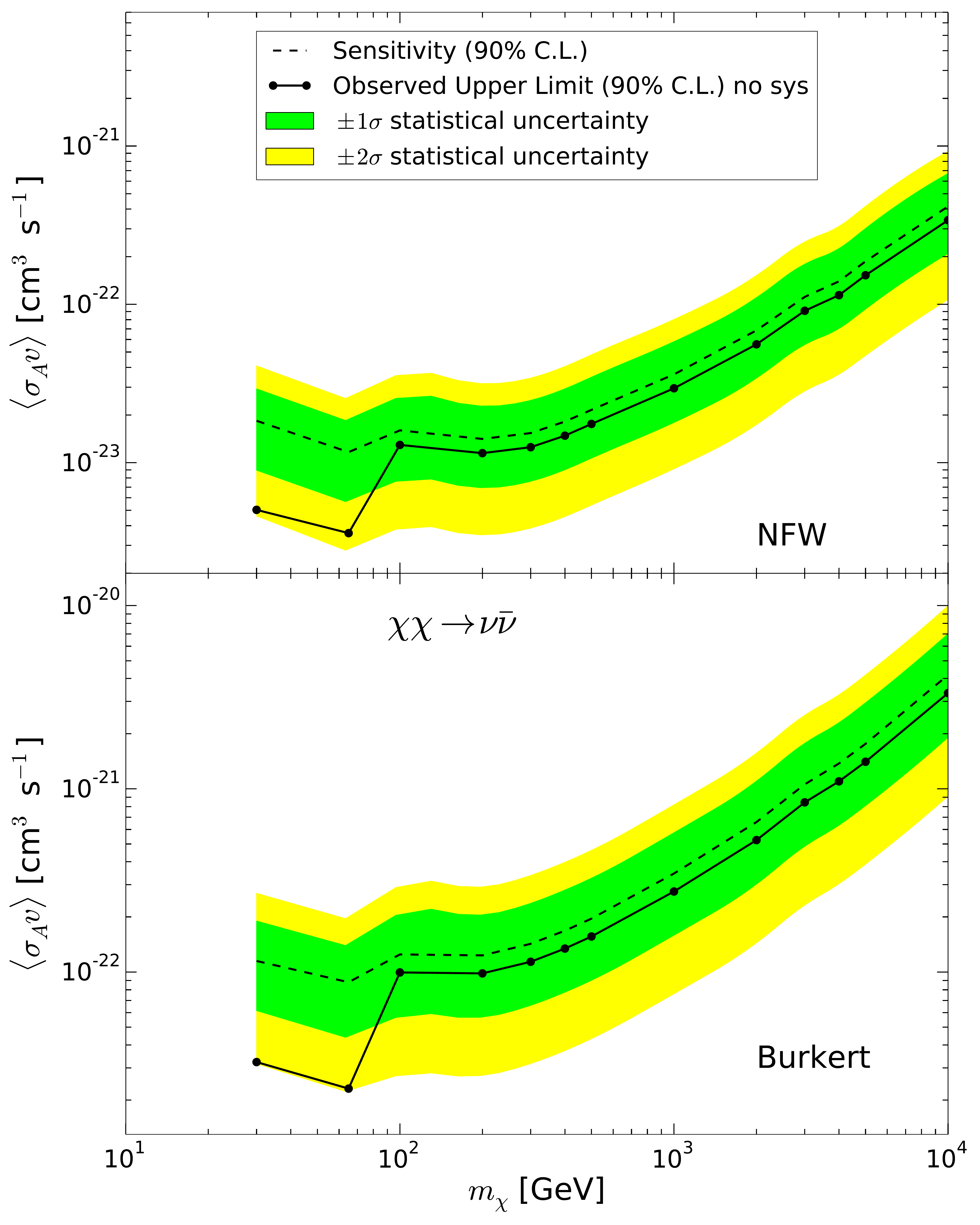}
    \caption[]{Sensitivity (dashed) and observed upper limit (solid, w/o systematics) at 90\% C.L. for WIMPs 
    annihilating to neutrinos assuming a NFW (top) and Burkert (bottom) DM halo profile.
    The statistical uncertainty on the sensitivity is shown at the $1\sigma$ (green band) and $2\sigma$ (yellow 
    band) level. The dip below 100\,GeV is caused by the under-fluctuation in the LE sample.
    }
    \label{fig:gclimNFWandBurkert}
\end{figure}

A small over-fluctuation was measured for the HE event selection.
However, after applying the likelihood analysis to both event selections
independently, all the resulting upper limits were smaller than their
corresponding sensitivity.
For the HE selection this implies that despite the over-fluctuation in
the number of events, the spatial distribution of these events within the
search window is incompatible with the expectation from dark matter
annihilation in the halo.
As an example, Fig.~\ref{fig:gclimNFWandBurkert} shows the sensitivity and the
observed upper limit after unblinding on $\langle \sigma_\mathrm{A} v \rangle$
for the NFW and Burkert profiles, assuming WIMP annihilation into neutrinos. In
addition the $\pm 1\sigma$ and $\pm 2\sigma$ statistical uncertainty bands of
the median upper limit are shown as green and yellow shaded areas,
respectively. The contribution from the LE and HE event selection can be
clearly seen through the upper limit curve (solid black line) being lower than
the expected median upper limit (dashed black line) in both cases, but to a
different extent, with the switch-over between $m_\chi = 100$~GeV and 200~GeV
for this annihilation channel.

\begin{table}[t]
	\caption{Results in terms of number of events for the two event selections. The differences of 
	observed and expected events in the on-source region, $\Delta n = n_\text{obs}-n_\text{bg}$, 
	show an under- and over-fluctuation of data for the LE and HE event selection, 
	respectively.}
	\label{tab:resultN}
	 \centering
	\begin{tabular}{ l| r r r}
		\hline
		    & $n_\text{obs}$ & $n_\text{bg}$ & $\Delta n$ \\
		\hline
		LE  	& $4,098$  & $4,217$  & $-119$ \\
		HE  	& $36,969$ & $36,806$ & $+162$ \\

	\end{tabular}
\end{table}

\begin{figure}[t]
    \centering
    \includegraphics[width=0.5\textwidth]{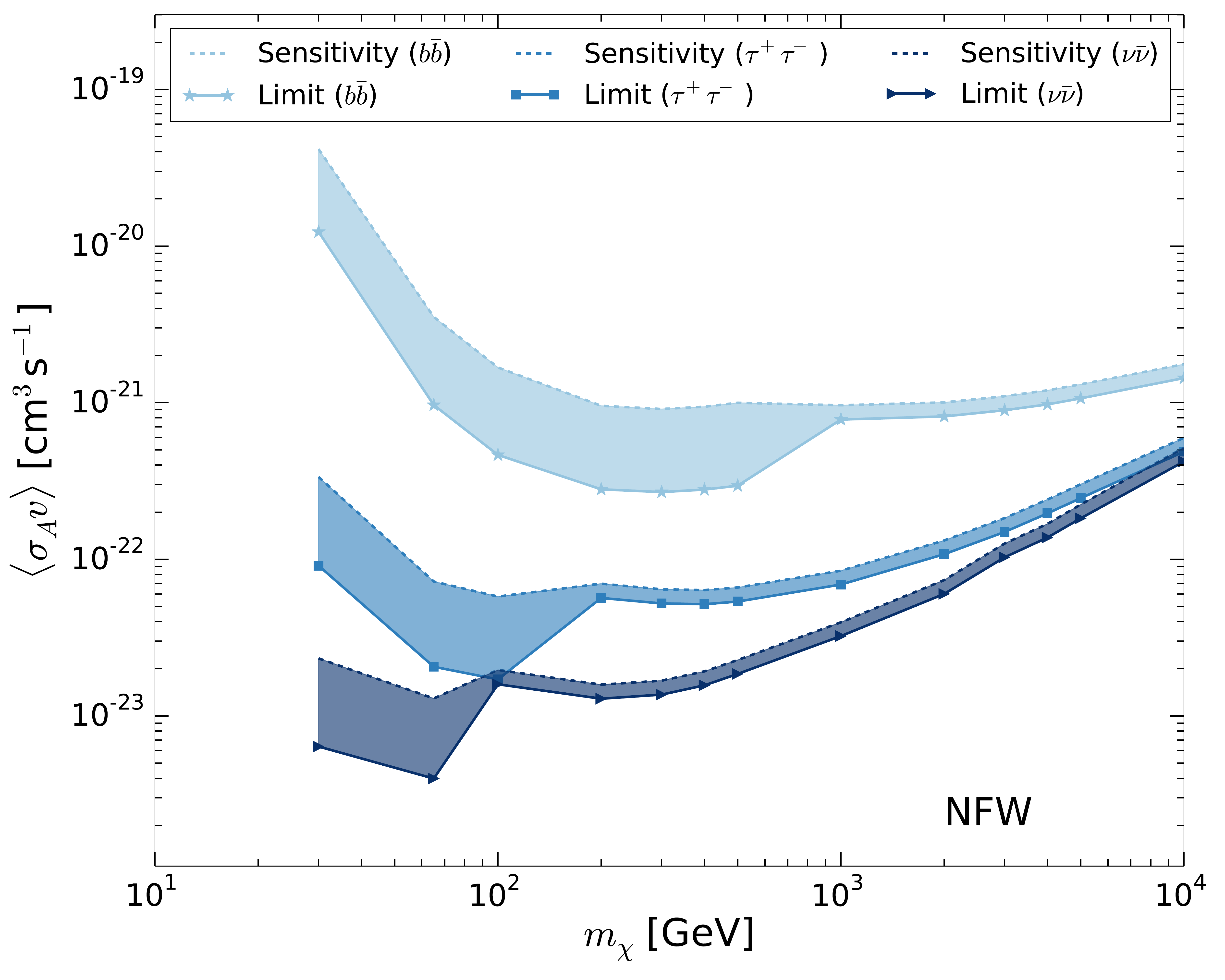}
    \caption[]{
    Sensitivity (dashed) and observed upper limits (solid) at 90\% C.L., including detector 
    systematics, for WIMPs annihilating to 
    $b\overline{b}$ (stars), $\tau^+\tau^-$ (squares), and directly to neutrinos (triangles) assuming a 
    NFW DM halo profile. The shaded areas are guides for the reader's eyes connecting sensitivity and limit 
    for a particular annihilation channel. 
    }
    \label{fig:comparison_sensitivities_limits}
\end{figure}

Table~\ref{tab:lim_table01} summarizes the upper limits for all considered 
annihilation channels and WIMP masses. The limits are shown separately for the two considered DM halo 
profiles; the NFW (top) and Burkert (bottom) profile.   
%
Figure~\ref{fig:comparison_sensitivities_limits} shows the sensitivity and limit
for three different annihilation channels. 

\begin{figure}[t]
    \centering
    \includegraphics[width=0.5\textwidth]{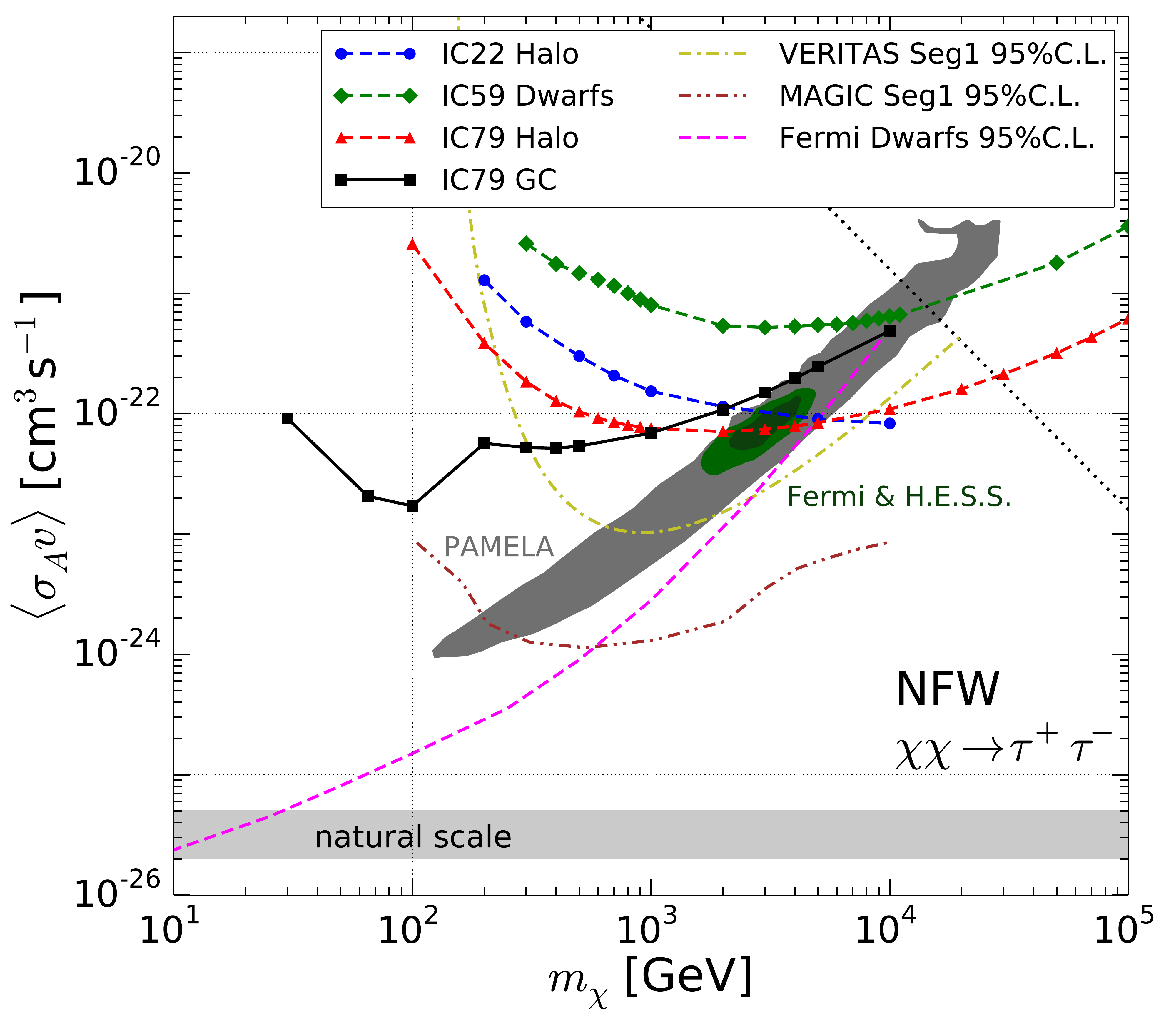}
    \caption[]{
    Comparison of limits from this work (IC79 GC) to other IceCube (designated by IC + ``string number'')
	searches for dark matter annihilation in self-bound structures 
\cite{2011PhRvD..84b2004A,Aartsen:2014hva,Aartsen:2013dxa}. Further, photon search limits
	from observation of dwarf spheroidals by VERITAS~\cite{2012PhRvD..85f2001A}, 
MAGIC~\cite{Aleksic:2013xea} and Fermi~\cite{Ackermann:2013yva} are shown.
    The grey-shaded region is a dark-matter interpretation of the positron
    excess reported by the PAMELA collaboration. The green-shaded regions are the
    $3\sigma$ and $5\sigma$ preferred regions from the $e^++e^-$-flux excess reported by
	Fermi and H.E.S.S. All shaded region data taken from~\cite{Meade:2009iu}. The region data
	and the IC22 halo limits are rescaled to the here assumed local dark matter density of
	$\rho_\mathrm{local}=0.471$\,GeV\,cm$^{-3}$. The natural scale is the
	self-annihilation cross-section region for WIMPs to be thermal relics from the Big 
	Bang~\cite{Steigman:2012nb}. The black dotted line in the upper right part
	of the figure is the unitarity bound~\cite{Griest:1989wd}. We note that preliminary Galactic Center 
limits from the ANTARES neutrino telescope have recently been released~\cite{Adrian-Martinez:2015wey}.
    }
    \label{fig:limit_comparison_other_exp}
\end{figure}

To compare the performance of this analysis to previous IceCube analyses and
other experiments, we choose the $\tau^+\tau^-$ annihilation channel assuming a
NFW DM halo profile. The comparison is shown in
Fig.~\ref{fig:limit_comparison_other_exp}. The black solid line shows the limit
of this analysis, whereas dashed lines with markers show the limits from previous galactic
halo~\cite{2011PhRvD..84b2004A,Aartsen:2014hva} and dwarf spheroidal galaxies
\cite{Aartsen:2013dxa} analyses with IceCube. The other lines show the limits from
gamma-ray experiments, in particular the limit from the dwarf spheroidal galaxy
Segue 1 analysis by VERITAS~\cite{2012PhRvD..85f2001A} (dash-dotted) and MAGIC~\cite{Aleksic:2013xea} 
(dash-dot-dotted), and the limit from
the Fermi analysis of several dwarf spheroidal
galaxies~\cite{Ackermann:2013yva} (dashed). Also shown is the DM interpretation of the
positron-fraction excess reported by the PAMELA collaboration (dark gray shaded
region) and the $3\sigma$ and $5\sigma$ preferred regions from the $e^+ +
e^-$-flux excess reported by the Fermi and H.E.S.S. collaborations as dark
green and green shaded regions, respectively. All the shaded region data are
taken from~\cite{Meade:2009iu} and rescaled to a local dark matter density of
$\rho_{\mathrm{local}}=0.471$\,GeV\,cm$^{-3}$ to this DM halo profile parameter
with the one considered in the other analyses. For a WIMP mass below 1TeV the
present analysis improves significantly in sensitivity on previous IceCube
analyses. Furthermore by using the DeepCore detector array, the
self-annihilation cross-section for WIMP masses below 100\,GeV is probed for
the first time by IceCube.

\input{input/table01}

\section{Conclusion}


We have presented limits on the cross-section on dark matter annihilation in the
Galactic Center, probing down to $\left< \sigma_\mathrm{A} v \right> \simeq
4\cdot10^{-24}\rm{cm}^3 \rm{s}^{-1}$ at 65\,GeV WIMP mass, assuming the NFW halo
profile and direct annihilation to neutrino pairs.

This analysis is the first IceCube Galactic Center DM search using the nearly complete
detector configuration. Further, it is the first IceCube DM search probing
$\left< \sigma_\mathrm{A} v \right>$ for WIMP masses below 100\,GeV by
utilizing the DeepCore infill-array of IceCube. 

We have presented methods for a selection of down-going muon neutrinos in
IceCube, making the southern hemisphere accessible to low-energy neutrino
searches in the energy range 10\,GeV~--~10\,TeV. These methods have been
applied to create two event selections, that are optimized for neutrino signals
from the direction of the Galactic Center.
Based on these event selections a likelihood analysis looking for a neutrino
flux from annihilating dark matter in the Galactic Center was performed, testing a
number of dark matter annihilation channels at different masses. 
The results are compatible with the background-only hypothesis, thus upper
limits on $\left< \sigma_\mathrm{A} v \right>$ were set (c.f.
Fig.~\ref{fig:limit_comparison_other_exp}). The limits from the low-energy
selection are almost $2\sigma$ lower than their sensitivity due to an
under-fluctuation in the number of background events. The limits presented here
for direct annihilation to $\nu\overline{\nu}$-pairs are model-independent and
conservative upper bounds for dark matter annihilation to Standard Model final
states~\cite{Beacom:2006tt}; even small branching ratios to other - more
visible - species at the $\left< \sigma_\mathrm{A} v \right>$-level presented
here would yield a detectable flux in gamma-ray experiments, or otherwise
stronger constraints. Thus, these limits complement gamma-ray detection
channels.

Future improvements to this analysis can be expected from improvements in the background
rejection in the energy region corresponding to the highest probed WIMP masses,
and the inclusion of an energy term in the likelihood function.

Long-term improvements should also be expected from possible IceCube
extensions.  The low-energy upgrade PINGU~\cite{Aartsen:2014oha} would increase
the sensitivity to low-mass WIMPs, and extend the probed mass range below
30\,GeV. PINGU is a possible future in-fill array with a denser instrumentation
than DeepCore. The high-mass (TeV-PeV) sensitivity would benefit from a future
high-energy extension, IceCube-Gen2~\cite{Aartsen:2014njl}. The aim for
IceCube-Gen2 is an expanded instrumented volume of the order of 10\,km$^3$ with
a larger inter-string spacing, compared to IceCube.

\input{input/ackn_18052015.tex}

\printbibliography

\end{document}

%% file: input/authors_25052015.tex
\author{IceCube Collaboration: M.~G.~Aartsen\thanksref{Adelaide}
\and K.~Abraham\thanksref{Munich}
\and M.~Ackermann\thanksref{Zeuthen}
\and J.~Adams\thanksref{Christchurch}
\and J.~A.~Aguilar\thanksref{BrusselsLibre}
\and M.~Ahlers\thanksref{MadisonPAC}
\and M.~Ahrens\thanksref{StockholmOKC}
\and D.~Altmann\thanksref{Erlangen}
\and T.~Anderson\thanksref{PennPhys}
\and M.~Archinger\thanksref{Mainz}
% ...
}
\thankstext{b}{Earthquake Research Institute, University of Tokyo, Bunkyo, Tokyo 113-0032, Japan}
\thankstext{c}{NASA Goddard Space Flight Center, Greenbelt, MD 20771, USA}
\institute{
% ...
School of Chemistry \& Physics, University of Adelaide, Adelaide SA, 5005 Australia \label{Adelaide}
% ...
\and Universit\'e Libre de Bruxelles, Science Faculty CP230, B-1050 Brussels, Belgium \label{BrusselsLibre}
% ...
\and Dept.~of Physics and Astronomy, University of Canterbury, Private Bag 4800, Christchurch, New Zealand \label{Christchurch}
% ...
\and Erlangen Centre for Astroparticle Physics, Friedrich-Alexander-Universit\"at Erlangen-N\"urnberg, D-91058 Erlangen, Germany \label{Erlangen}
% ...
\and Dept.~of Physics and Wisconsin IceCube Particle Astrophysics Center, University of Wisconsin, Madison, WI 53706, USA \label{MadisonPAC}
\and Institute of Physics, University of Mainz, Staudinger Weg 7, D-55099 Mainz, Germany \label{Mainz}
% ...
\and Technische Universit\"at M\"unchen, D-85748 Garching, Germany \label{Munich}
% ...
\and Oskar Klein Centre and Dept.~of Physics, Stockholm University, SE-10691 Stockholm, Sweden \label{StockholmOKC}
% ...
\and Dept.~of Physics, Pennsylvania State University, University Park, PA 16802, USA \label{PennPhys}
% ...
\and DESY, D-15735 Zeuthen, Germany \label{Zeuthen}
} 

%% file: input/table01.tex
\begin{Large}\begin{table}
        \caption{Final upper limits (including detector systematics) on the self-annihilation
        cross-section, $\langle \sigma_{A} v \rangle$,
        for different annihilation channels and WIMP masses, $m_\chi$, for the NFW (top) and Burkert
        (bottom) DM halo profiles.}
        \label{tab:lim_table01} 
        \centering
    \resizebox{\columnwidth}{!}{
\begin{tabular}{r|c|c|c|c|c}\hline
$m_\chi \,[\mathrm{GeV}]$ & \multicolumn{5}{c}{ $\langle \sigma_A v\rangle \left[ 10^{-22} \mathrm{cm}^{3}\mathrm{s}^{-1} \right]$ assuming NFW profile} \\
&$b \bar {b}$&$W^+W^-$&$\tau^+\tau^-$&$\mu^+\mu^-$&$\nu\bar{\nu}$\\
\hline 
$30$ &\mbox{\makebox[5mm][r]{$120$}\makebox[1mm][c]{.}\makebox[5mm][l]{$0$}}&---&\mbox{\makebox[5mm][r]{$0$}\makebox[1mm][c]{.}\makebox[5mm][l]{$91$}}&\mbox{\makebox[5mm][r]{$0$}\makebox[1mm][c]{.}\makebox[5mm][l]{$78$}}&\mbox{\makebox[5mm][r]{$0$}\makebox[1mm][c]{.}\makebox[5mm][l]{$064$}}\\
$65$ &\mbox{\makebox[5mm][r]{$9$}\makebox[1mm][c]{.}\makebox[5mm][l]{$7$}}&---&\mbox{\makebox[5mm][r]{$0$}\makebox[1mm][c]{.}\makebox[5mm][l]{$21$}}&\mbox{\makebox[5mm][r]{$0$}\makebox[1mm][c]{.}\makebox[5mm][l]{$17$}}&\mbox{\makebox[5mm][r]{$0$}\makebox[1mm][c]{.}\makebox[5mm][l]{$04$}}\\
$100$ &\mbox{\makebox[5mm][r]{$4$}\makebox[1mm][c]{.}\makebox[5mm][l]{$6$}}&\mbox{\makebox[5mm][r]{$0$}\makebox[1mm][c]{.}\makebox[5mm][l]{$35$}}&\mbox{\makebox[5mm][r]{$0$}\makebox[1mm][c]{.}\makebox[5mm][l]{$17$}}&\mbox{\makebox[5mm][r]{$0$}\makebox[1mm][c]{.}\makebox[5mm][l]{$14$}}&\mbox{\makebox[5mm][r]{$0$}\makebox[1mm][c]{.}\makebox[5mm][l]{$16$}}\\
$200$ &\mbox{\makebox[5mm][r]{$2$}\makebox[1mm][c]{.}\makebox[5mm][l]{$8$}}&\mbox{\makebox[5mm][r]{$1$}\makebox[1mm][c]{.}\makebox[5mm][l]{$1$}}&\mbox{\makebox[5mm][r]{$0$}\makebox[1mm][c]{.}\makebox[5mm][l]{$57$}}&\mbox{\makebox[5mm][r]{$0$}\makebox[1mm][c]{.}\makebox[5mm][l]{$49$}}&\mbox{\makebox[5mm][r]{$0$}\makebox[1mm][c]{.}\makebox[5mm][l]{$13$}}\\
$300$ &\mbox{\makebox[5mm][r]{$2$}\makebox[1mm][c]{.}\makebox[5mm][l]{$7$}}&\mbox{\makebox[5mm][r]{$1$}\makebox[1mm][c]{.}\makebox[5mm][l]{$0$}}&\mbox{\makebox[5mm][r]{$0$}\makebox[1mm][c]{.}\makebox[5mm][l]{$52$}}&\mbox{\makebox[5mm][r]{$0$}\makebox[1mm][c]{.}\makebox[5mm][l]{$46$}}&\mbox{\makebox[5mm][r]{$0$}\makebox[1mm][c]{.}\makebox[5mm][l]{$14$}}\\
$400$ &\mbox{\makebox[5mm][r]{$2$}\makebox[1mm][c]{.}\makebox[5mm][l]{$8$}}&\mbox{\makebox[5mm][r]{$1$}\makebox[1mm][c]{.}\makebox[5mm][l]{$1$}}&\mbox{\makebox[5mm][r]{$0$}\makebox[1mm][c]{.}\makebox[5mm][l]{$52$}}&\mbox{\makebox[5mm][r]{$0$}\makebox[1mm][c]{.}\makebox[5mm][l]{$46$}}&\mbox{\makebox[5mm][r]{$0$}\makebox[1mm][c]{.}\makebox[5mm][l]{$16$}}\\
$500$ &\mbox{\makebox[5mm][r]{$2$}\makebox[1mm][c]{.}\makebox[5mm][l]{$9$}}&\mbox{\makebox[5mm][r]{$1$}\makebox[1mm][c]{.}\makebox[5mm][l]{$1$}}&\mbox{\makebox[5mm][r]{$0$}\makebox[1mm][c]{.}\makebox[5mm][l]{$54$}}&\mbox{\makebox[5mm][r]{$0$}\makebox[1mm][c]{.}\makebox[5mm][l]{$48$}}&\mbox{\makebox[5mm][r]{$0$}\makebox[1mm][c]{.}\makebox[5mm][l]{$19$}}\\
$1000$ &\mbox{\makebox[5mm][r]{$7$}\makebox[1mm][c]{.}\makebox[5mm][l]{$8$}}&\mbox{\makebox[5mm][r]{$1$}\makebox[1mm][c]{.}\makebox[5mm][l]{$5$}}&\mbox{\makebox[5mm][r]{$0$}\makebox[1mm][c]{.}\makebox[5mm][l]{$69$}}&\mbox{\makebox[5mm][r]{$0$}\makebox[1mm][c]{.}\makebox[5mm][l]{$63$}}&\mbox{\makebox[5mm][r]{$0$}\makebox[1mm][c]{.}\makebox[5mm][l]{$32$}}\\
$2000$ &\mbox{\makebox[5mm][r]{$8$}\makebox[1mm][c]{.}\makebox[5mm][l]{$2$}}&\mbox{\makebox[5mm][r]{$2$}\makebox[1mm][c]{.}\makebox[5mm][l]{$3$}}&\mbox{\makebox[5mm][r]{$1$}\makebox[1mm][c]{.}\makebox[5mm][l]{$1$}}&\mbox{\makebox[5mm][r]{$1$}\makebox[1mm][c]{.}\makebox[5mm][l]{$0$}}&\mbox{\makebox[5mm][r]{$0$}\makebox[1mm][c]{.}\makebox[5mm][l]{$6$}}\\
$3000$ &\mbox{\makebox[5mm][r]{$8$}\makebox[1mm][c]{.}\makebox[5mm][l]{$9$}}&\mbox{\makebox[5mm][r]{$3$}\makebox[1mm][c]{.}\makebox[5mm][l]{$1$}}&\mbox{\makebox[5mm][r]{$1$}\makebox[1mm][c]{.}\makebox[5mm][l]{$5$}}&\mbox{\makebox[5mm][r]{$1$}\makebox[1mm][c]{.}\makebox[5mm][l]{$5$}}&\mbox{\makebox[5mm][r]{$1$}\makebox[1mm][c]{.}\makebox[5mm][l]{$0$}}\\
$4000$ &\mbox{\makebox[5mm][r]{$9$}\makebox[1mm][c]{.}\makebox[5mm][l]{$7$}}&\mbox{\makebox[5mm][r]{$3$}\makebox[1mm][c]{.}\makebox[5mm][l]{$9$}}&\mbox{\makebox[5mm][r]{$2$}\makebox[1mm][c]{.}\makebox[5mm][l]{$0$}}&\mbox{\makebox[5mm][r]{$2$}\makebox[1mm][c]{.}\makebox[5mm][l]{$0$}}&\mbox{\makebox[5mm][r]{$1$}\makebox[1mm][c]{.}\makebox[5mm][l]{$4$}}\\
$5000$ &\mbox{\makebox[5mm][r]{$11$}\makebox[1mm][c]{.}\makebox[5mm][l]{$0$}}&\mbox{\makebox[5mm][r]{$4$}\makebox[1mm][c]{.}\makebox[5mm][l]{$8$}}&\mbox{\makebox[5mm][r]{$2$}\makebox[1mm][c]{.}\makebox[5mm][l]{$5$}}&\mbox{\makebox[5mm][r]{$2$}\makebox[1mm][c]{.}\makebox[5mm][l]{$5$}}&\mbox{\makebox[5mm][r]{$1$}\makebox[1mm][c]{.}\makebox[5mm][l]{$8$}}\\
$10000$ &\mbox{\makebox[5mm][r]{$14$}\makebox[1mm][c]{.}\makebox[5mm][l]{$0$}}&\mbox{\makebox[5mm][r]{$8$}\makebox[1mm][c]{.}\makebox[5mm][l]{$4$}}&\mbox{\makebox[5mm][r]{$4$}\makebox[1mm][c]{.}\makebox[5mm][l]{$9$}}&\mbox{\makebox[5mm][r]{$5$}\makebox[1mm][c]{.}\makebox[5mm][l]{$4$}}&\mbox{\makebox[5mm][r]{$4$}\makebox[1mm][c]{.}\makebox[5mm][l]{$2$}}\\
\hline
\hline
$m_\chi \,[\mathrm{GeV}]$ & \multicolumn{5}{c}{ $\langle \sigma_A v\rangle \left[ 10^{-22} \mathrm{cm}^{3}\mathrm{s}^{-1} \right]$ assuming Burkert profile} \\
&$b \bar {b}$&$W^+W^-$&$\tau^+\tau^-$&$\mu^+\mu^-$&$\nu\bar{\nu}$\\
\hline 
$30$ &\mbox{\makebox[5mm][r]{$4400$}\makebox[1mm][c]{.}\makebox[5mm][l]{$0$}}&---&\mbox{\makebox[5mm][r]{$5$}\makebox[1mm][c]{.}\makebox[5mm][l]{$6$}}&\mbox{\makebox[5mm][r]{$4$}\makebox[1mm][c]{.}\makebox[5mm][l]{$9$}}&\mbox{\makebox[5mm][r]{$0$}\makebox[1mm][c]{.}\makebox[5mm][l]{$41$}}\\
$65$ &\mbox{\makebox[5mm][r]{$61$}\makebox[1mm][c]{.}\makebox[5mm][l]{$0$}}&---&\mbox{\makebox[5mm][r]{$1$}\makebox[1mm][c]{.}\makebox[5mm][l]{$3$}}&\mbox{\makebox[5mm][r]{$1$}\makebox[1mm][c]{.}\makebox[5mm][l]{$1$}}&\mbox{\makebox[5mm][r]{$0$}\makebox[1mm][c]{.}\makebox[5mm][l]{$26$}}\\
$100$ &\mbox{\makebox[5mm][r]{$30$}\makebox[1mm][c]{.}\makebox[5mm][l]{$0$}}&\mbox{\makebox[5mm][r]{$3$}\makebox[1mm][c]{.}\makebox[5mm][l]{$3$}}&\mbox{\makebox[5mm][r]{$1$}\makebox[1mm][c]{.}\makebox[5mm][l]{$1$}}&\mbox{\makebox[5mm][r]{$0$}\makebox[1mm][c]{.}\makebox[5mm][l]{$91$}}&\mbox{\makebox[5mm][r]{$1$}\makebox[1mm][c]{.}\makebox[5mm][l]{$2$}}\\
$200$ &\mbox{\makebox[5mm][r]{$18$}\makebox[1mm][c]{.}\makebox[5mm][l]{$0$}}&\mbox{\makebox[5mm][r]{$8$}\makebox[1mm][c]{.}\makebox[5mm][l]{$9$}}&\mbox{\makebox[5mm][r]{$4$}\makebox[1mm][c]{.}\makebox[5mm][l]{$3$}}&\mbox{\makebox[5mm][r]{$3$}\makebox[1mm][c]{.}\makebox[5mm][l]{$8$}}&\mbox{\makebox[5mm][r]{$1$}\makebox[1mm][c]{.}\makebox[5mm][l]{$1$}}\\
$300$ &\mbox{\makebox[5mm][r]{$17$}\makebox[1mm][c]{.}\makebox[5mm][l]{$0$}}&\mbox{\makebox[5mm][r]{$8$}\makebox[1mm][c]{.}\makebox[5mm][l]{$6$}}&\mbox{\makebox[5mm][r]{$4$}\makebox[1mm][c]{.}\makebox[5mm][l]{$2$}}&\mbox{\makebox[5mm][r]{$3$}\makebox[1mm][c]{.}\makebox[5mm][l]{$8$}}&\mbox{\makebox[5mm][r]{$1$}\makebox[1mm][c]{.}\makebox[5mm][l]{$3$}}\\
$400$ &\mbox{\makebox[5mm][r]{$18$}\makebox[1mm][c]{.}\makebox[5mm][l]{$0$}}&\mbox{\makebox[5mm][r]{$9$}\makebox[1mm][c]{.}\makebox[5mm][l]{$2$}}&\mbox{\makebox[5mm][r]{$4$}\makebox[1mm][c]{.}\makebox[5mm][l]{$4$}}&\mbox{\makebox[5mm][r]{$3$}\makebox[1mm][c]{.}\makebox[5mm][l]{$9$}}&\mbox{\makebox[5mm][r]{$1$}\makebox[1mm][c]{.}\makebox[5mm][l]{$4$}}\\
$500$ &\mbox{\makebox[5mm][r]{$19$}\makebox[1mm][c]{.}\makebox[5mm][l]{$0$}}&\mbox{\makebox[5mm][r]{$10$}\makebox[1mm][c]{.}\makebox[5mm][l]{$0$}}&\mbox{\makebox[5mm][r]{$4$}\makebox[1mm][c]{.}\makebox[5mm][l]{$7$}}&\mbox{\makebox[5mm][r]{$4$}\makebox[1mm][c]{.}\makebox[5mm][l]{$2$}}&\mbox{\makebox[5mm][r]{$1$}\makebox[1mm][c]{.}\makebox[5mm][l]{$7$}}\\
$1000$ &\mbox{\makebox[5mm][r]{$60$}\makebox[1mm][c]{.}\makebox[5mm][l]{$0$}}&\mbox{\makebox[5mm][r]{$13$}\makebox[1mm][c]{.}\makebox[5mm][l]{$0$}}&\mbox{\makebox[5mm][r]{$6$}\makebox[1mm][c]{.}\makebox[5mm][l]{$3$}}&\mbox{\makebox[5mm][r]{$5$}\makebox[1mm][c]{.}\makebox[5mm][l]{$8$}}&\mbox{\makebox[5mm][r]{$3$}\makebox[1mm][c]{.}\makebox[5mm][l]{$0$}}\\
$2000$ &\mbox{\makebox[5mm][r]{$67$}\makebox[1mm][c]{.}\makebox[5mm][l]{$0$}}&\mbox{\makebox[5mm][r]{$21$}\makebox[1mm][c]{.}\makebox[5mm][l]{$0$}}&\mbox{\makebox[5mm][r]{$10$}\makebox[1mm][c]{.}\makebox[5mm][l]{$0$}}&\mbox{\makebox[5mm][r]{$9$}\makebox[1mm][c]{.}\makebox[5mm][l]{$7$}}&\mbox{\makebox[5mm][r]{$5$}\makebox[1mm][c]{.}\makebox[5mm][l]{$8$}}\\
$3000$ &\mbox{\makebox[5mm][r]{$75$}\makebox[1mm][c]{.}\makebox[5mm][l]{$0$}}&\mbox{\makebox[5mm][r]{$28$}\makebox[1mm][c]{.}\makebox[5mm][l]{$0$}}&\mbox{\makebox[5mm][r]{$14$}\makebox[1mm][c]{.}\makebox[5mm][l]{$0$}}&\mbox{\makebox[5mm][r]{$14$}\makebox[1mm][c]{.}\makebox[5mm][l]{$0$}}&\mbox{\makebox[5mm][r]{$9$}\makebox[1mm][c]{.}\makebox[5mm][l]{$8$}}\\
$4000$ &\mbox{\makebox[5mm][r]{$84$}\makebox[1mm][c]{.}\makebox[5mm][l]{$0$}}&\mbox{\makebox[5mm][r]{$36$}\makebox[1mm][c]{.}\makebox[5mm][l]{$0$}}&\mbox{\makebox[5mm][r]{$18$}\makebox[1mm][c]{.}\makebox[5mm][l]{$0$}}&\mbox{\makebox[5mm][r]{$18$}\makebox[1mm][c]{.}\makebox[5mm][l]{$0$}}&\mbox{\makebox[5mm][r]{$12$}\makebox[1mm][c]{.}\makebox[5mm][l]{$0$}}\\
$5000$ &\mbox{\makebox[5mm][r]{$92$}\makebox[1mm][c]{.}\makebox[5mm][l]{$0$}}&\mbox{\makebox[5mm][r]{$44$}\makebox[1mm][c]{.}\makebox[5mm][l]{$0$}}&\mbox{\makebox[5mm][r]{$23$}\makebox[1mm][c]{.}\makebox[5mm][l]{$0$}}&\mbox{\makebox[5mm][r]{$23$}\makebox[1mm][c]{.}\makebox[5mm][l]{$0$}}&\mbox{\makebox[5mm][r]{$17$}\makebox[1mm][c]{.}\makebox[5mm][l]{$0$}}\\
$10000$ &\mbox{\makebox[5mm][r]{$130$}\makebox[1mm][c]{.}\makebox[5mm][l]{$0$}}&\mbox{\makebox[5mm][r]{$76$}\makebox[1mm][c]{.}\makebox[5mm][l]{$0$}}&\mbox{\makebox[5mm][r]{$45$}\makebox[1mm][c]{.}\makebox[5mm][l]{$0$}}&\mbox{\makebox[5mm][r]{$50$}\makebox[1mm][c]{.}\makebox[5mm][l]{$0$}}&\mbox{\makebox[5mm][r]{$41$}\makebox[1mm][c]{.}\makebox[5mm][l]{$0$}}\\
\hline
\end{tabular}
}
\end{table}
\end{Large}

%% file: input/ackn_18052015.tex
\begin{acknowledgements}

We acknowledge the support from the following agencies:
U.S. National Science Foundation-Office of Polar Programs,
U.S. National Science Foundation-Physics Division,
University of Wisconsin Alumni Research Foundation,
the Grid Laboratory Of Wisconsin (GLOW) grid infrastructure at the University of Wisconsin - Madison, the Open Science Grid (OSG) grid infrastructure;
U.S. Department of Energy, and National Energy Research Scientific Computing Center,
the Louisiana Optical Network Initiative (LONI) grid computing resources;
Natural Sciences and Engineering Research Council of Canada,
WestGrid and Compute/Calcul Canada;
Swedish Research Council,
Swedish Polar Research Secretariat,
Swedish National Infrastructure for Computing (SNIC),
and Knut and Alice Wallenberg Foundation, Sweden;
German Ministry for Education and Research (BMBF),
Deutsche Forschungsgemeinschaft (DFG),
Helmholtz Alliance for Astroparticle Physics (HAP),
Research Department of Plasmas with Complex Interactions (Bochum), Germany;
Fund for Scientific Research (FNRS-FWO),
FWO Odysseus programme,
Flanders Institute to encourage scientific and technological research in industry (IWT),
Belgian Federal Science Policy Office (Belspo);
University of Oxford, United Kingdom;
Marsden Fund, New Zealand;
Australian Research Council;
Japan Society for Promotion of Science (JSPS);
the Swiss National Science Foundation (SNSF), Switzerland;
National Research Foundation of Korea (NRF);
Danish National Research Foundation, Denmark (DNRF)

\end{acknowledgements}